# Doming and spin cascade in Ferric Haems: Femtosecond X-ray Absorption and X-ray Emission Studies

*Camila Bacellar[1], Dominik Kinschel[1], Giulia F. Mancini[1], Rebecca A. Ingle[1], Jérémy Rouxel[1], Oliviero Cannelli[1], Claudio Cirelli[2], Gregor Knopp[2], Jakub Szlachetko[3], Frederico A. Lima[4], Samuel Menzi[2], Georgios Pamfilidis[2], Katharina Kubicek[4], Dmitry Khakhulin[4], Wojciech Gawelda[4,5], Angel Rodriguez-Fernandez[4], Mykola Biednov[4], Christian Bressler[4], Christopher A. Arrell[2], Philip J. M. Johnson[2], Christopher Milne[2] and Majed Chergui[1]\**

[1] Ecole Polytechnique Fédérale de Lausanne (EPFL), Laboratoire de Spectroscopie Ultrarapide, ISIC and Lausanne Centre for Ultrafast Science (LACUS), CH-1015 Lausanne, Switzerland

[2] SwissFEL, Paul-Scherrer-Institut (PSI), 5232 Villigen PSI, Switzerland

[3] Institute of Nuclear Physics, Polish Academy of Sciences, 31-342 Kraków, Poland

[4] European XFEL, Holzkoppel 4, D-22869, Schenefeld, Germany

[5] Faculty of Physics, Adam Mickiewicz University, ul. Uniwersytetu Poznanskiego 2, 61-614 Poznan, Poland

\* Majed.chergui@epfl.ch

**Abstract**

**The structure-function relationship is at the heart of biology and major protein deformations are correlated to specific functions. In the case of heme proteins, doming is associated with the respiratory function in hemoglobin and myoglobin, while ruffling has been correlated with electron transfer processes, such as in the case of Cytochrome c (Cyt c). The latter has indeed evolved to become an important electron transfer protein in humans. In its ferrous form, it undergoes ligand release and doming upon photoexcitation, but its ferric form does not release the distal ligand, while the return to the ground state has been attributed to thermal relaxation. Here, by combining femtosecond Fe K-edge X-ray absorption near-edge structure (XANES) studies and femtosecond Fe K$_\alpha$ and K$_\beta$ X-ray emission spectroscopy (XES), we demonstrate that the photocycle of ferric Cyt c is entirely due to a cascade among excited spin states of the Iron ion, causing the ferric heme to undergo doming, which we identify for the first time. We also argue that this pattern is common to all ferric haems, raising the question of the biological relevance of doming in such proteins.**

**Introduction:**

Cytochrome c is a small protein that mediates the electron transfer (ET) from cytochrome c reductase to cytochrome c oxidase.[1] It also plays a role in apoptosis and its conformationally dependent peroxidase activity.[2-4] Cyt c consists of an iron proto-porphyrin IX complex, with the central Fe atom hexacoordinated by the four N pyrrole atoms (Np) of the porphyrin, a distal methionine ligand (Met80) and a proximal histidine (His18) ligand (Figure 1). The latter anchors the haem to the protein peptide chain and is also linked to the protein via two thioether covalent bonds with Cysteine-14 and -17. Cyt c Several studies have been conducted aimed at relating the ET properties of Cyt c to structural factors that govern, e.g., the reduction potentials and the electronic coupling between donor and acceptor.[5-7] The Cyt c haem is characterised by a large ruffling distortion (an out-of-plane distortion of the porphyrin, see figure S1b) induced by the protein fold and the haem motif. Haem ruffling is the dominant out-of-plane (OOP) deformation in c-type cytochromes involved in ET[8-10] and in nitrophorins involved in NO transport.[11-13] These OOP distortions are energetically unfavourable,[14] but as they have been conserved through evolution, it was hypothesized that ruffling is crucial for tuning the reduction potential of the protein and therefore its ET properties.[15-17] NMR studies[15] showed that ruffling destabilizes all three occupied Fe 3d-based molecular orbitals (Figure S1b), and it was hypothesized that this controls the redox potential.

Another major haem deformation that has been conserved through evolution is doming (Figure S1c). It has been well characterised and correlated to functions such as oxygen storage and transport in haemoglobins (Hb's) and myoglobins (Mb's).[18-19] Doming results from the occupation of anti-bonding Fe d-orbitals, i.e. formation of higher metal spin states (Figure S1c). In the case of ferrous hemoproteins (Hb, Mb or Cyt c), upon distal ligand release from the Fe atom, the pentacoordinated haem adopts a domed deoxyMb configuration in a high spin (HS) quintet state (Figure S1c). This, and the reverse process (from domed to planar), are considered the initial molecular-level events of the respiratory function in Hb.[18]

Ever since it was discovered that ligand detachment from ferrous haems can be induced with high quantum yield by absorption of ultraviolet-visible light,[20] photoexcitation has been used to mimic the basic events of the respiratory function, i.e. the ligand detachment from, and recombination to the Fe atom. This was particularly the case with the advent of ultrafast spectroscopy as ferrous haem proteins were among the first systems ever to be investigated by such a method.[21] By photoexciting the haem porphyrin either into its Q- (visible) or Soret-

(ultraviolet, UV) bands, or even into higher lying states, the distal ligand detachment-recombination events in ferrous haems have been monitored using various optical (UV, visible, infrared, Raman) probes.[22-25] It was concluded that prompt photodissociation of the ligand from the low spin planar haem occurs, accompanied by a simultaneous doming of the haem porphyrin, i.e. formation of the high spin (HS, quintet) deoxyMb. Evidence of ligand dissociation and doming are based on three types of observables: a) the transient UV-visible absorption (TA) spectra in the region of the Q- and Soret-bands reproduce the difference between the steady-state UV-visible absorption spectra of the HS penta-coordinated domed deoxyMb and the LS hexacoordinated ferrous planar ligated haems;[26] b) the transient resonance Raman studies, which exhibit the shift of the Fe-His bond frequency from its value in the planar ligated haem ($\sim$265 cm$^{-1}$) to its value in the deoxyMb form ($\sim$220 cm$^{-1}$);[23, 27] c) ligand dissociation is also witnessed by the appearance of the infrared bands of the unbound diatomic ligand in Carboxymyoglobin (MbCO),[28-30] Nitrosylmyoglobin (MbNO)[31] and oxymyoglobin (MbO$_2$).[29, 32]

In the case of ferric haem proteins such as metmyoglobin (metMb),[26] Cyt c,[23] Cyanomyoglobin (MbCN),[33] azidomyoglobin (MbN$_3$),[34] and the protein sensor oxy-FixLH,[35] either one or more of the above observables for haem doming and/or ligand release were missing, which lead to the conclusion that ligand dissociation and doming do not take place in ferric haems. Specifically for ferric Cyt c, the description has prevailed that the electronically photoexcited haem decays to high lying vibrational levels of the ground state, which then undergoes thermal relaxation.[7, 23, 25, 36-37] It was further concluded[24] that the latter causes haem photoreduction,[6, 24, 27] but the origin of the reducing electron is unclear, although nearby amino-acid residues, such as Tyrosine 67 (Figure 1), have been proposed as possible candidates.[5]

The scenario of a pure thermal relaxation in ferric haems was recently challenged by ultrafast infrared (IR) to UV-visible TA studies on ferric MbCN,[33] metMb[38] and MbN$_3$,[34] suggesting that a cascade through excited spin states may be involved, though not ruling out a partial parallel thermal relaxation pathway. These states lie between the LUMO (Q-state) and HOMO (ground state) of the haem porphyrin and are due to population of metal d-orbitals (Figure S3).[39] They cannot be optically accessed because of selection rules. In fact, UV-visible probes only monitor the HOMO-LUMO (e.g. the Q- and Soret-bands) of the porphyrin ring, and they are not sensitive to spin states of the metal. Therefore, the above conclusion that relaxation of the photoexcited haem proceeds via excited spin states[33-34, 38] was not based on their direct observation, but was inferred from the relaxation kinetics and, whenever this was the case, on IR spectral features that show up upon photoexcitation.[33-34]

X-ray absorption (XAS) and emission spectroscopy (XES) can circumvent the limitations of optical spectroscopies as they are element-selective and structure- and spin-sensitive.[40-41] In recent years, time-domain (femtoseconds to picoseconds) X-ray spectroscopies have emerged as powerful tools to probe the electronic structure changes and the structural dynamics of (bio)molecular and of materials with the additional advantage that they can detect optically-silent transient states.[42] Quasi-steady state[43-45] and fs-ps Fe K-edge XANES have already been used to probe the photoinduced electronic and structural changes in photoexcited ferrous hemoproteins, such as MbCO,[46-47] MbNO,[48] and Cyt c[49] upon distal ligand dissociation. Similar studies were also performed at the Co K-edge on Cyanocobalamin (Vitamin $B_{12}$) using polarised XAS.[50-51] All these studies concluded the ligand dissociation and confirmed formation of a domed porphyrin. Furthermore, XES has emerged as a reliable marker of the spin state of the transition metal atom (number of unpaired 3d electrons) via the $K_\beta$ lines (3p→1s emission) that reflect the 3p-3d exchange interactions and exhibit an intensity decrease and blue energy shift of the $K_\beta$ line with increasing spin, while the $K_{\beta'}$ sideband increases in intensity as can be seen in the spectra of reference compounds in Figures S2b and c.[40] The 2p-3d interactions are weaker than the 3p-3d interactions, yet $K_\alpha$ XES is also a marker of spin via the linear dependence of the $K_{\alpha 1}$ line full-width at half-maximum (FWHM) and the exchange interaction of the core hole with the number of unpaired spins in the valence shell, up to $S=3/2$.[52] Femtosecond Fe $K_\beta$ XES studies revealed the details of the ultrafast spin cross-over (SCO) in photoexcited $[Fe(bpy)_3]^{2+}$,[53] and was also used to identify the high spin (HS, S=2) deoxy haem product of photoexcited ferrous Cytochrome c after dissociation of its distal methionine ligand.[49] In the present work, we specifically address the relaxation cascade in photoexcited ferric mitochondrial horse heart Cyt c using element-specific and structure- and spin-sensitive methods such as fs Fe K-edge X-ray near-edge absorption structure (XANES) and fs Fe $K_\alpha$ and $K_\beta$ X-ray emission spectroscopy (XES, Figure S2a) at the Swiss Free Electron Laser (SwissFEL, Villigen) and the European XFEL (Eu-XFEL, Hamburg). We show that the decay of the photoexcited haem in ferric Cyt c entirely proceeds via high spin states of the metal, which we identify by both their structural XANES and their electronic XES signatures. Details of the experimental procedures are given in the SI.

**Results and Discussion:**

Ferric Cyt c has a low spin (LS, S=1/2) ground state with 5 electrons in the Fe-$d_{xy}$, $d_{xz}$, $d_{yz}$ ($t_{2g}$) orbitals (Figure S3). The intermediate spin (IS, S=3/2) and HS (S=5/2) can be formed when one and two electrons, respectively occupy the $d_{z2}$ and $d_{x2-y2}$ antibonding orbitals. In principle, this will lead to elongation of the Fe-$N_p$ bonds and therefore doming. The latter is expected to be larger for S=5/2 (two electrons in the antibonding orbitals) than for S=3/2 (only one electron in the antibonding orbitals).[54] Figure 2 shows the steady-state Fe K-edge XANES of ferric Cyt c (black trace), which agrees with the literature (Figure S4),[55] along with the transient spectrum (laser-on minus laser-off spectra) 500 fs time delay after 350 nm excitation. The transient shows an increased absorption at the edge energy (~7125 eV) followed by a negative signal between 7130 and 7160 eV, and a positive one beyond. The intensity increase at the edge can either be caused by a reduction of the Fe atom,[56] an increase of Fe-N bond distances (as reported for [Fe(bpy)$_3$]$^{2+}$ [57]), or a combination of both. We rule out photoreduction of the Fe atom based on the fact that its yield is low (<2% for 403 nm excitation)[24, 27] and its rise time is slow (~5 ps).[24] Further support to this conclusion is the comparison of our transient with the difference of the steady-state ferrous minus ferric Cyt c spectra, which strongly deviate from each other (Figure S5). This type of comparison is commonly used to identify oxidation state changes.[41] On the other hand, the transient in figure 2 resembles those reported in quasi-steady state XANES studies of MbCO under cw irradiation[58-59] and ps XANES studies of photoexcited ferrous MbCO,[46] MbNO,[48] and more recently, in fs studies of ferrous Cyt c.[49] In all these cases, the oxidation state of the Fe atom did not change and the transient features were caused by the fact that after detachment of the distal ligand from the planar LS hexacoordinated haem, the pentacoordinated haem undergoes doming, i.e. the Fe-$N_p$ bonds elongate. This is clearly established by the correspondence between the XANES transients with the difference of steady-state domed deoxyMb minus planar ligated Mb XANES spectra.[46, 48]

In order to highlight the resemblance of the ferric and ferrous haem transients, in Figures S6 to S8 we compare the ferric Cyt c transient with those for MbNO, MbCO and ferrous Cyt c. To this purpose, the present ferric transient had to be shifted by ~-2 eV to best overlap the ferrous one.[49] This amount corresponds to the Fe K-edge shift of the ferric and ferrous Cyt c species.[55] Thus apart from this shift, which further confirms that no reduction has taken place, the ferric Cyt c transient strongly resembles those of ferrous species, in particular Cyt c. Some differences also show up, e.g. the shoulder at ~7125 eV in ferrous Cyt c, or the larger width of the main feature in MbNO, the level of agreement is satisfactory and on the level of agreement between the various ferrous species. In the latter case, and even though the photoproduct is the same (deoxyMb), the differences between transients originates from the differences in the ground state XANES spectra.[60] Thus, whether the system is ferric or ferrous and regardless of a detached ligand or lack thereof (assumed so far for ferric Cyt c), the main effect on the profile of the transient is that a domed photoproduct is formed. One could conclude from figures S6 to S8 that the domed species in ferric Cyt c may also be pentacoordinated. While this has been excluded based on transient optical and resonance Raman studies,[23, 26] the ps XANES studies on MbNO, which forms a domed ligated haem upon NO recombination, showed only a minor effect of the distal ligand on the shape of the XANES transients.[48] This is also the case when one examines the simulated XANES of ferrous Cyt c for different distances between the distal ligand and the Fe atom.[49] Thus, while the present data cannot exclude or support formation of a pentacoordinated haem, we rely on the conclusions of the optical studies that the ligand does not dissociate.[25,28] However, we conclude that doming via formation of HS states does occur in a ferric haem. This is further supported by a study of the temperature dependence of the Fe K-edge XANES of the hydroxide complex of ferric Mb ($Mb^{III}OH^-$) between 80 and 300 K[61] for which prior magnetic susceptibilities[62] had established that it is purely LS at 80 K and predominantly HS at 300 K. In their study, Oyanagi et al[61] found that the peak at ~7.125 keV in their difference spectra (high temperature minus 80 K spectrum) clearly increases with temperature and reproduces the same trends as the magnetic susceptibility measurements. From the above considerations, it is fair to conclude that the effect determining the overall shape of the transients in Figures 2 and S6 to S8 is doming, via population of an IS or HS state.

Further to the XANES structural signature, the electronic signature of the IS/HS state is provided by the transient $K_\alpha$ and $K_\beta$ XES spectra. Because $K_\beta$ XES is well-established to identify spin states (Figure S2),[63] we show in figure 3 the $K_\beta$ transient at 200 fs time delay. It suffers from a poor signal-to-noise ratio because it rides on a high background of elastically scattered light. Nevertheless, its shape is clearly reminiscent of the signature of the HS (S=2) state reported for $[Fe(bpy)_3]^{2+}$[53] and ferrous Cyt c,[49] whose transient at 600 fs is also shown (red trace). In comparing the transients, the ferrous ones were shifted by 1-2 eV to the red in order to better overlap the ferric transient. This further supports the above observation of a lack of photoreduction at short times. The characteristic features (Figure S2b,c) of a HS state, i.e. the blue shift and intensity decrease of the $K_\beta$ line around 7058 eV, and the increase of the $K_{\beta'}$ sideband at 7043 eV, are clearly reproduced. Furthermore, we also show in this figure the difference $K_\beta$ steady-state spectra of reference sextet (S=5/2) and quartet (S=3/2) minus doublet (S=1/2) porphyrins.[64] Both reproduce the experimental transient well, though they do not allow the identification of the spin of the state populated at 200 fs time delay. Nevertheless, there is no doubt that a state with a higher spin than the ground state is formed.

$K_\alpha$ XES is one order of magnitude more intense than the $K_\beta$ emission and further away from the elastic peak, suffering less from the inelastic background signal. Therefore, we anticipate a better signal-to-noise ratio, and this is indeed the case. Laser-off (t<0) and laser-on (t=200 fs) $K_\alpha$ spectra are shown in figure 4, along with the corresponding transient (blue circles). A more complete set of laser-on spectra and their transients is shown in figure S9, and figure S10 (left panels) zooms into the evolution of the laser-on $K_{\alpha 1}$ line. It can be seen that its broadening is largest at the earliest time delays, and it decreases with time thereafter. This broadening is more pronounced on the blue wing than on the red (Figure S10) as expected for a spin increase.[40, 52] In the absence of reference $K_\alpha$ spectra of ferric haem LS, IS and HS states and in order to quantify the broadening, we fitted the laser-off and laser-on spectra using an asymmetric pseudo-Voigt line-shape (described in § S6),[65] which best reproduces the emission profile (Figure S10 top). The results of the fit (figure S10, right panels) are shown separately from the actual experimental line shapes (Figure S10, left panels) for the sake of clarity. However, as an example, Figure S11 shows the result of the fit of the $K_\alpha$ transient at 200 fs: the difference of the fitted laser-off and laser-on $K_{\alpha 1}$ lines is compared with the actual transient showing very good agreement above 6403 eV. In fitting the laser-on $K_{\alpha 1}$ line, a broadening of ~0.7 eV was retrieved, which is of the order of magnitude expected for a transition from a LS to a HS state.[40, 52] Since it was reported that the $K_\alpha$ line-shape does not significantly change with spin beyond S=3/2,[52] we can only conclude that the final state has $S \geq 3/2$.

In summary, both the XANES signal, via a structural signature, and the $K_\alpha$ and $K_\beta$ XES signals via an electronic signature point to formation of IS or HS states within the first ps. Further insight into the relaxation dynamics is provided by the temporal traces of the signals, which are shown in Figure 5a for the XANES signal at 7.125 keV (maximum amplitude) and figure 5b for the FWHM of the fit of laser-on $K_{\alpha 1}$ lines. Both traces exhibit a prompt rise, within the instrument response function IRF≈140±30 fs for the XANES and ~150 fs for the XES (see § S2), followed by a decay that is best fit using a bi-exponential function convoluted to the Gaussian IRF (see § S5). They both yield consistent time constants of ~600 fs and ~8 ps, in very good agreement with previous optical TA studies (Table 1).[25] Considering that the XES is a pure electronic signature and is not sensitive to vibrational effects, we can conclude that these times reflect an electronic relaxation among spin states. However, the pre-exponential factors differ between the XANES and the XES signals (table 1). This is due to the fact that the XANES transient signal at 7.125 keV reflects a structural signature, whose intensity is not simply linear with the Fe-N bond elongation (see e.g. [57]). Although the latter is largest for the S=5/2 state, the S=3/2 state surely contributes to the signal but is not distinguishable in the XANES transient. From the kinetics, we can conclude that the ~600 fs time scale is due to decay of the S=3/2 state, while the ~8 ps is due to the S=5/2 state relaxing back to the ground state (S=1/2). Further supporting this conclusion is that in figure S9 (bottom) the positive peak at 6406 eV and the negative one at 6404 eV of the $K_{\alpha 1}$ transient show somewhat different evolutions. By integrating the signal of the positive and negative peaks and plotting them as a function of time (Figure S12a), we recover two different kinetic traces (figure S12b). Despite the few data points, it can be seen that the negative part shows a prompt rise with a biexponential decay, comparable to Figure 5, while the positive signal shows a slower rise and a mono-exponential decay. Fitting these kinetic traces with multiexponential functions yields time constants comparable to those derived from Figure 5. In particular, the short decaying component (~700 fs) of the negative signal is quite close to the rise of the positive signal, suggesting a cascade process in which an intermediate state feeds population into the HS state. This analysis also reveals hitherto unknown spin information contained in the $K_{\alpha}$ spectra, which deserves further investigation for a more general class of systems.

Finally, comparing the time constants derived here with those from optical experiments (table 1) provides a complete picture of the photocycle of ferric Cyt c. The optical fluorescence monitors the rise and decay of the porphyrin LUMO,[66] which both occur at ultrafast time scales (<50 fs), and the decay was attributed to relaxation to metal spin states. The optical TA experiments also probe the porphyrin π−π* transitions, but they monitor the recovery of the ground state (HOMO) for which time constants of ~700 fs, ~3.5 ps and ~11 ps had been retrieved.[25] On the other hand, the XANES and XES observables selectively sense the electronic distribution on the Fe atom. Thus, the prompt rise of the X-ray time traces reflects rapid (<50 fs) formation of an excited metal spin state, most likely the S=3/2 state, via decay of the LUMO[66] and with no change of the metal oxidation state. The subsequent decay of the X-ray signal occurs on two timescales (~600 fs and ~8 ps), which are identical to the first and third components reported in the optical TA (Table 1). Our results show that these times are not due to a hot haem vibrational relaxation as previously claimed[7, 23, 25, 37] (XES is not sensitive to thermal effects), but rather to a population relaxation through spin states. The X-ray kinetic traces can thus be explained by a cascade via a promptly formed S=3/2 state that relaxes in 600-700 fs to the S=5/2 state, which then takes ~8 ps to return to the initial S=1/2 ground state, as schematically shown in Figure 6. Of course, a parallel vibrational cooling channel in the return to the initial state is also possible, but the close correspondence between the time constants of the present study and the optical ones (table 1), supports the conclusion that the ~600 fs and ~8 ps time scales are due to an electronic relaxation cascade. In this respect, the intermediate (~3.5 ps) component reported in most optical studies[7, 23, 25] is likely due to a thermal relaxation within the HS state. We also performed a tri-exponential function to fit the X-ray time traces (figure 5) and did retrieve an additional ~3 ps intermediate component but its weight was weak and uncertainty large, while the first and third components changed neither in decay times nor in relative weights. We therefore hold the bi-exponential fit of the X-ray data as physically more meaningful. Further support that the relaxation is predominantly electronic, i.e. with minimal parallel thermal relaxation channels comes from the estimates of the photolysis yield from the XANES transient signal compared to the photoexcitation yield based on the experimental parameters (energy/pulse, focal spot, concentration, absorption coefficient, sample thickness, etc.) presented in § S4. These estimates agree to within a factor of 2, which bearing in mind that scattering and reflection losses of the laser pump beam are not considered, confirm that nearly all photoexcited species relax by an electronic cascade via spin states.

The relaxation scheme depicted in figure 6 goes beyond the case of ferric Cyt c and can be generalised to other ferric haemoproteins. Indeed, it supports previous conclusions from optical studies that spin states are involved in the relaxation of photoexcited ferric MbCN,[33] MbN$_3$,[34] and metMb.[38] This is further reflected by the rather similar kinetic behaviour of all these systems upon excitation of either the Q- or Soret-bands (table S1). Three time-scales emerge, sub-ps, ps and few ps, which vary somewhat from system to system due to the different distal ligands but show a good overall agreement. Thus, the evidence is compelling that haem relaxation proceeds via a cascade through spin states leading to haem doming without ligand dissociation in all the thus-far investigated ferric haems. However, the intermediate $\tau_2 \approx 3$ ps time scale in table S1 reflects vibrational relaxation in the HS state and is similar to the value reported for various haem proteins both ferrous and ferric,[23, 36, 67] as well as [Fe(bpy)$_3$]$^{2+}$.[68-69]

Based on the present and previous results, doming is a deformation that is present in all haem proteins, whose role in the case of ferrous Mb and Hb is well-established in the respiratory function. The question arises regarding its biological significance in ferric haems, as functionally diverse as Cyt c, metMb, MbCN, MbN$_3$, MbOH and oxy-FixLH. Doming, i.e. HS states, may play a role if their energies with respect to the LS states is at kT. In this respect, we note that contrary to its ferrous counterpart that exhibits a high thermal stability, requiring temperatures above 100 °C (373 K) to undergo unfolding,[70] ferric Cyt c at pH7 starts unfolding at considerably lower temperatures of 40 °C (313 K) to 70 °C (343 K), going through different unfolding states.[71-73] While resonance Raman spectra indicate that the LS state of the haem iron is maintained in some of these states,[74] it is unclear to what extent the remaining conformational transitions are related to higher spin states and whether these have a functionally relevant role. Further studies are needed to clarify the biological role of doming in ferric haems. Quite remarkably in ferric $MbN_3$, it was found that the photocycle is identical whether one uses excitation of the Q-band in the green or excitation of the $N_3$ stretch mode in the region of 2000 cm$^{-1}$, and the occurrence of a HS state was invoked.[34] This suggests the HS state is low lying, in line with the above discussed results on MbOH.[61] On the other hand, for the ferrous carboxyhaemoglobin (HbCO) infrared driven vibrational climbing up to the v=6 level of the CO stretch did not lead to ligand dissociation, while visible excitation leads to a unity quantum yield ligand dissociation.[75] While this may be due to the way the ligand redistributes its vibrational energy into the haem, these different behaviours underscore the importance of identifying the exact energy of the HS state with respect to the LS state in both ferric and ferrous haems.

**Conclusions:**

In summary, in the present study we have combined element-specific structure-sensitive XANES with spin-sensitive XES to study the ferric form of cytochrome c. We have shown that its photocycle is entirely due to a cascade among spin states, that occurs via a prompt decay of the LUMO into the S=3/2 intermediate spin state, followed by a further decay in ~600-700 fs into the high spin S=5/2 state, which causes doming. The S=5/2 state decays back to the ground S=1/2 state in ~10 ps. Except for the time scales which differ, this process is identical to the spin cross-over dynamics in Fe-polypyridine complexes.[76-77] The occurrence of doming in haem proteins is not feature limited to the ferrous forms, but it also includes ferric ones. The fact that doming seems to be present in very diverse ferric haems raises the question of its biological role.

**Acknowledgment**: This work was supported by the Swiss NSF via the NCCR:MUST and grants 200020_169914 and 200021_175649 and the European Research Council Advanced Grants H2020 ERCEA 695197 DYNAMOX. W.G. acknowledges partial support from the National Science Center (NCN) in Poland under SONATA BIS 6 grant No. 2016/22/E/ST4/00543. CB and GM benefited from the InterMUST Women Postdoc

Fellowships. We acknowledge swissFEL (PSI-Villigen, Switzerland) and European XFEL (Schenefeld, Germany) for provision of X-ray free-electron laser beamtime at the Alvra and FXE instruments, respectively, as well as for their staff for support.

**Data availability statement**

Raw data were generated at SwissFEL and the European XFEL large-scale facilities. Derived data supporting the findings of this study are available from the corresponding author upon request.

**Table 1:** Comparison of the rise ($\tau_r$) and decay times ($\tau_i$) of the transient XANES reported in this work with previous ultrafast optical studies carried by fluorescence up-conversion spectroscopy[66] and UV-visible transient absorption (TA) spectroscopy.[25] The optical fluorescence monitors the rise and the decay of the LUMO of the porphyrin macrocycle (i.e. the Q-bands), while the TA monitors the return of the population to the HOMO ground state. Both methods are sensitive to $\pi$–$\pi^*$ transitions of the porphyrin. The X-ray signal is sensitive to the electron density on the Fe atom and therefore it monitors its redistribution after photoexcitation. The $a_i$'s are the pre-exponential factors. The error bars are based on confidence intervals of the fit. All time constants are in picoseconds.

| Method (pump wavelength) | $\tau_r$ | $\tau_1$ ($a_1$) | $\tau_2$ ($a_2$) | $\tau_3$ ($a_3$) |
|---|---|---|---|---|
| XANES (350 nm), this work | 0.14±0.03 | 0.61±0.2 (0.70) | -- | 8.7±4 (0.30) |
| K$_\alpha$ XES (400 nm), this work | 0.15±0.03 | 0.63±0.12 (0.50) | -- | 7.8±4 (0.50) |
| Ultrafast Fluorescence of the LUMO (400 nm)[66] | ≤ 0.04 | ≤ 0.04 | | |
| Optical TA (530 nm)[25] | 0.22±0.02 | 0.70±0.09 | 3.4±0.3 | 11±2 |

**Figure Captions**

Figure 1: Crystal structure of ferric Cytochrome C (Cyt c). The haem, and the methionine 80 (Met80), Histidine 18 (His18) and Tyrosine 67 (Tyr67) amino-acid residues ligated to the Fe haem are highlighted as sticks (Fe [orange], C [teal], N [blue], O [red], S [yellow]). PDB: 1HRC. The Fe atom at the centre of the porphyrin is coordinated by four pyrrole ($N_p$) atoms and by the distal Met80 and the proximal His18 ligands.

Figure 2: Steady state XANES ground state spectrum of ferric cytochrome c in room temperature aqueous solutions at pH 7 (black trace) and transient spectrum (excited minus unexcited) recorded 500 fs time delay after femtosecond excitation at 350 nm (Brown dots).

Figure 3: Normalized transient $K_\beta$ XES of ferric Cyt c recorded at 200 fs (grey dots) compared to the steady-state difference $K_\beta$ XES spectra of sextet (Fe(III)(TPP)Cl) minus doublet (Fe(III)(bpy)$_3$, purple), and quartet (Fe(III)(PTC)Cl) minus doublet (Fe(III)(bpy)$_3$, green) from refs [53, 64] (TPP=tetraphenylporphyrin, PTC=Phtalocyanine Chloride). We also compare our results with the digitized transient spectrum of ferrous Cyt c at 600 fs time delay (red) from ref. [49]. Note that the latter and the steady-state difference spectra were red shifted by ~1-2 eV to match the present data points.

Figure 4: Steady-state (laser-off) Fe-$K_\alpha$ emission spectrum of ferric Cytochrome c in room temperature aqueous solutions at pH7, showing the $K_{\alpha 1}$ (6404 eV) and $K_{\alpha 2}$ (6391 eV) lines (black trace). The laser-on spectrum at 200 fs time delay (green trace) and transient XES (laser-on minus laser-off spectrum) spectrum at 200 fs time delay after excitation at 400 nm (blue dots).

Figure 5: Temporal evolution of the transient XANES and XES signals. a) intensity of the XANES signal at 7125.3 eV. b) FWHM extracted from an asymmetric Voigt fit of the $K_{\alpha 1}$ line at 6405 keV (see § S6 and figure S12). The black trace represents the fit using a function with an IRF-limited rise (~140 fs for the XAS and ~150 fs for the XES) and a bi-exponential decay. The fit parameters (time constants and pre-exponential factors) are given in table S1.

Figure 6: Schematic representation of the relaxation cascade following UV-visible excitation of ferric Cytochrome C. The energy levels are arbitrarily placed. Excitation of the system leads to an ultrafast population and decay of the Q-state (the porphyrin LUMO) as established by fluorescence up-conversion studies,[78] the intermediate spin S=3/2 state is promptly populated and decays in ~650 fs to the high spin S=5/2 state, which then decays back to the ground state in ~10 ps.

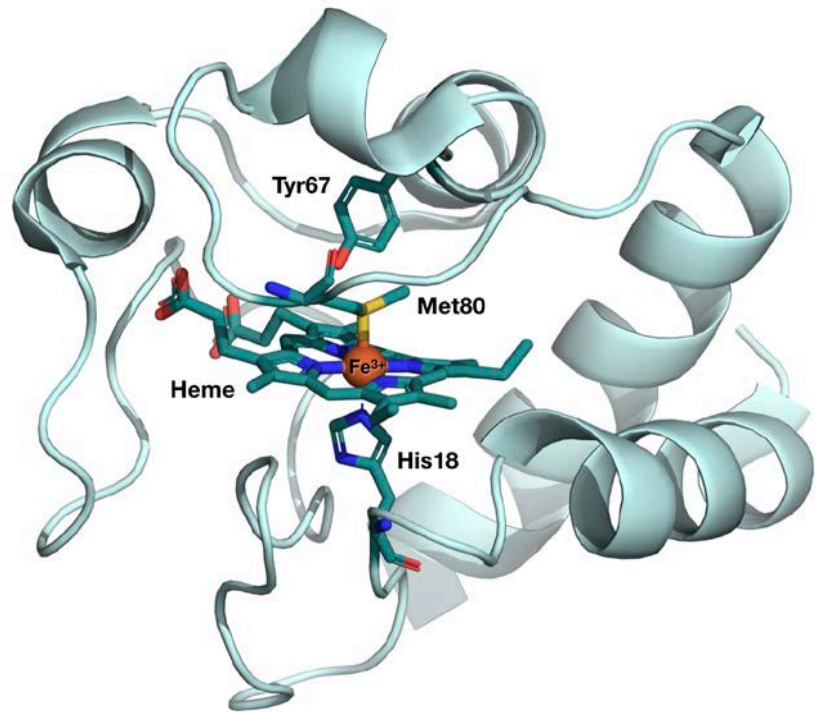

**Figure 1**

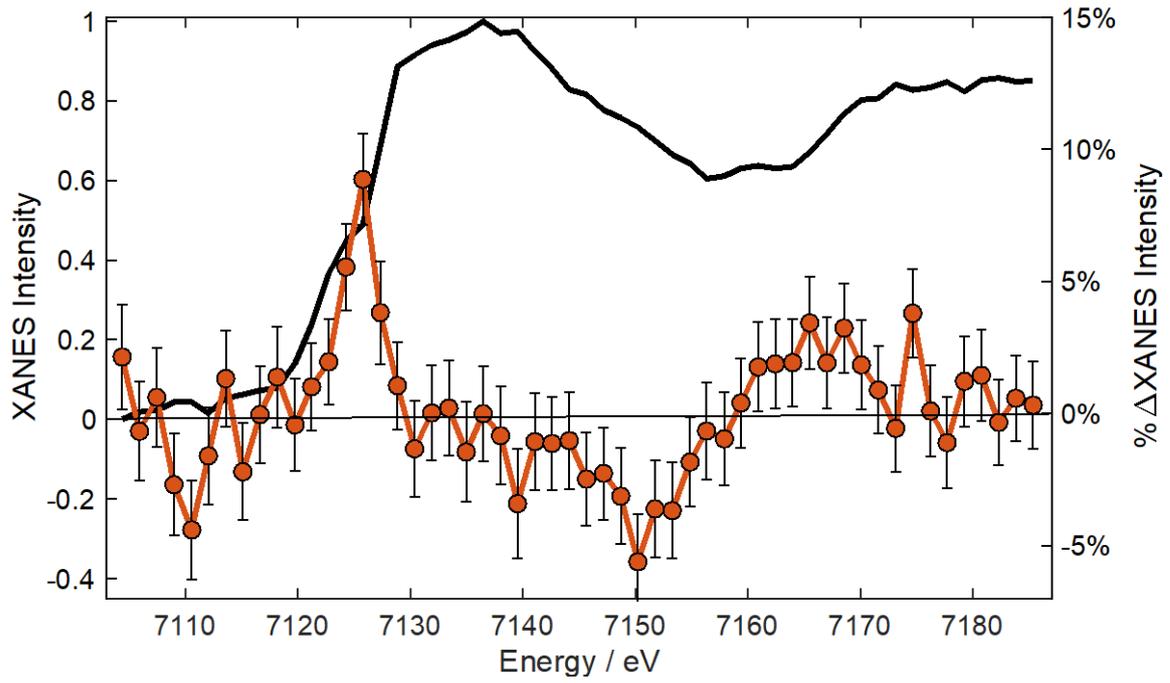

**Figure 2**

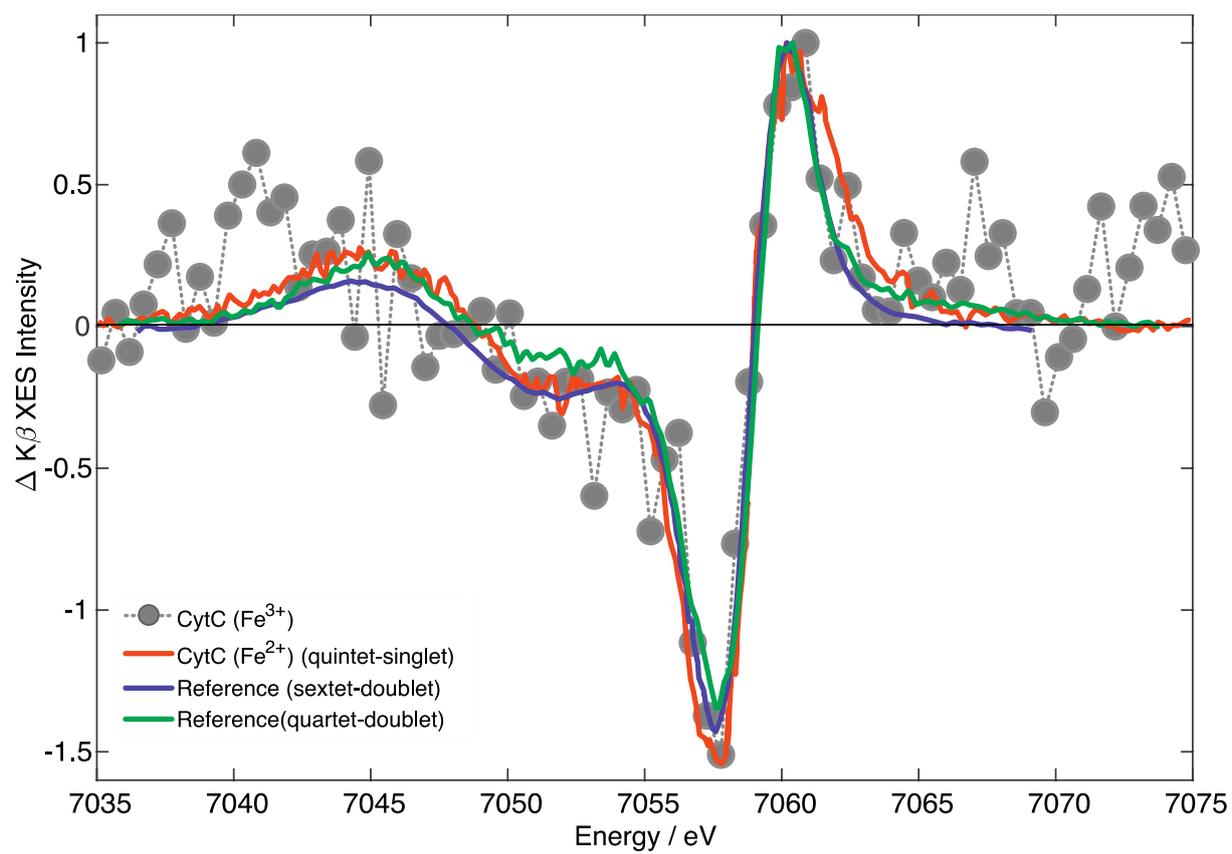

Figure 3

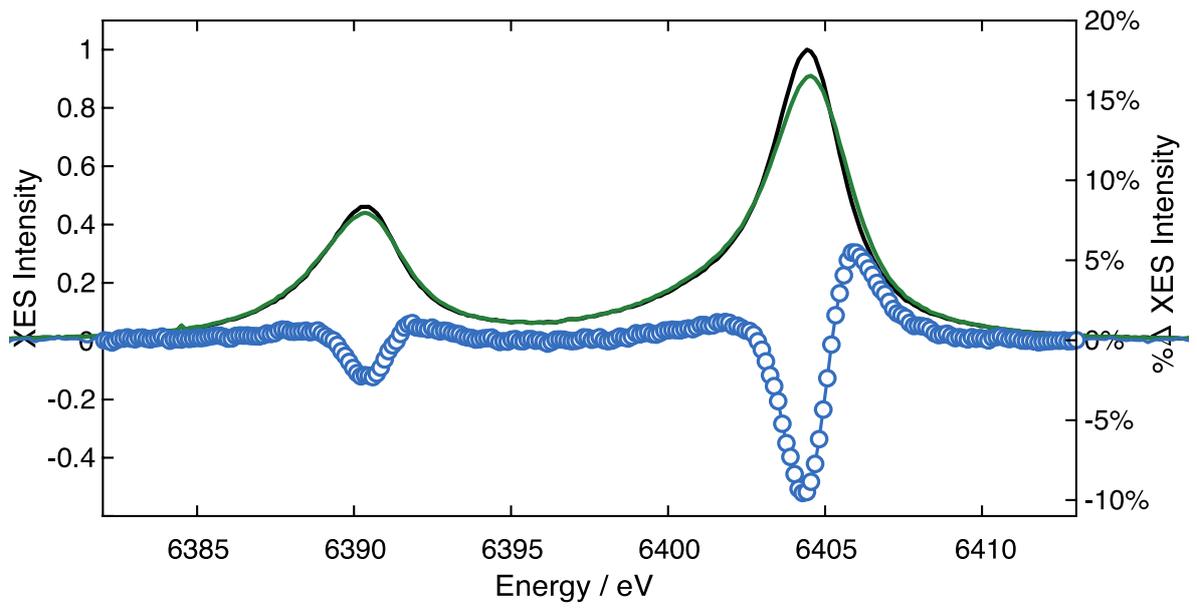

**Figure 4**

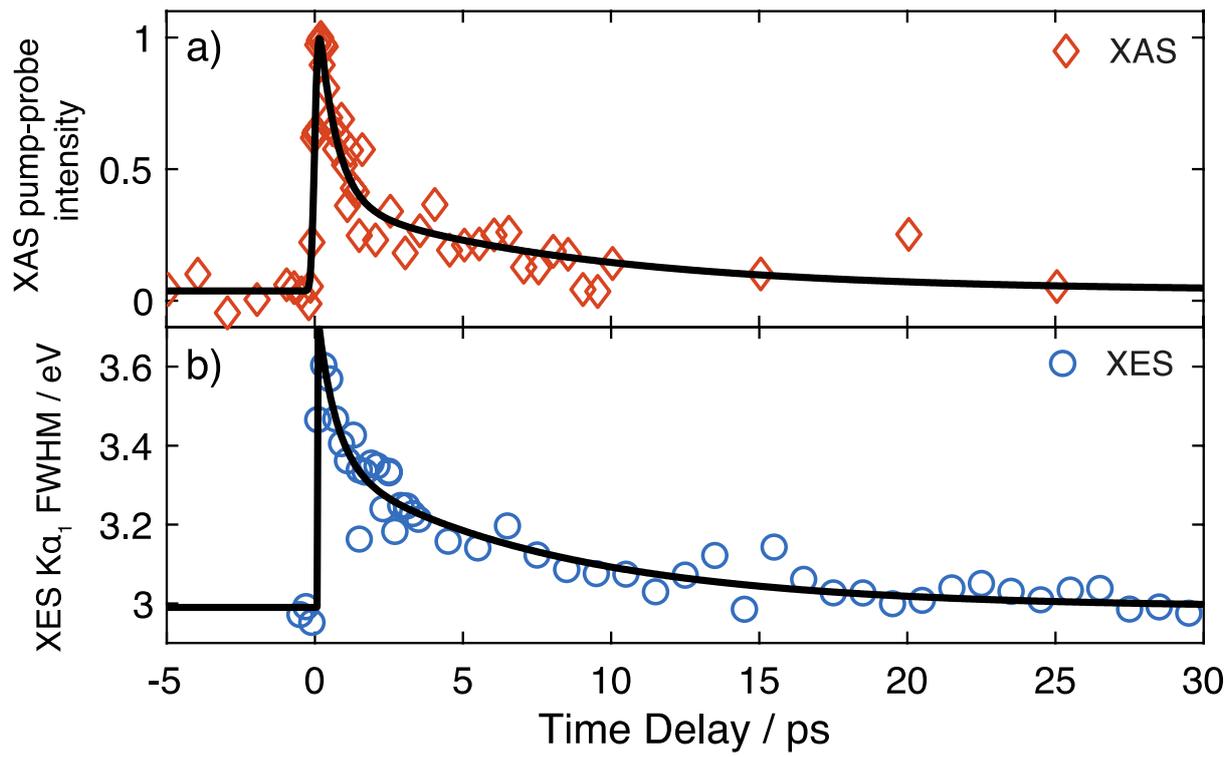

**Figure 5**

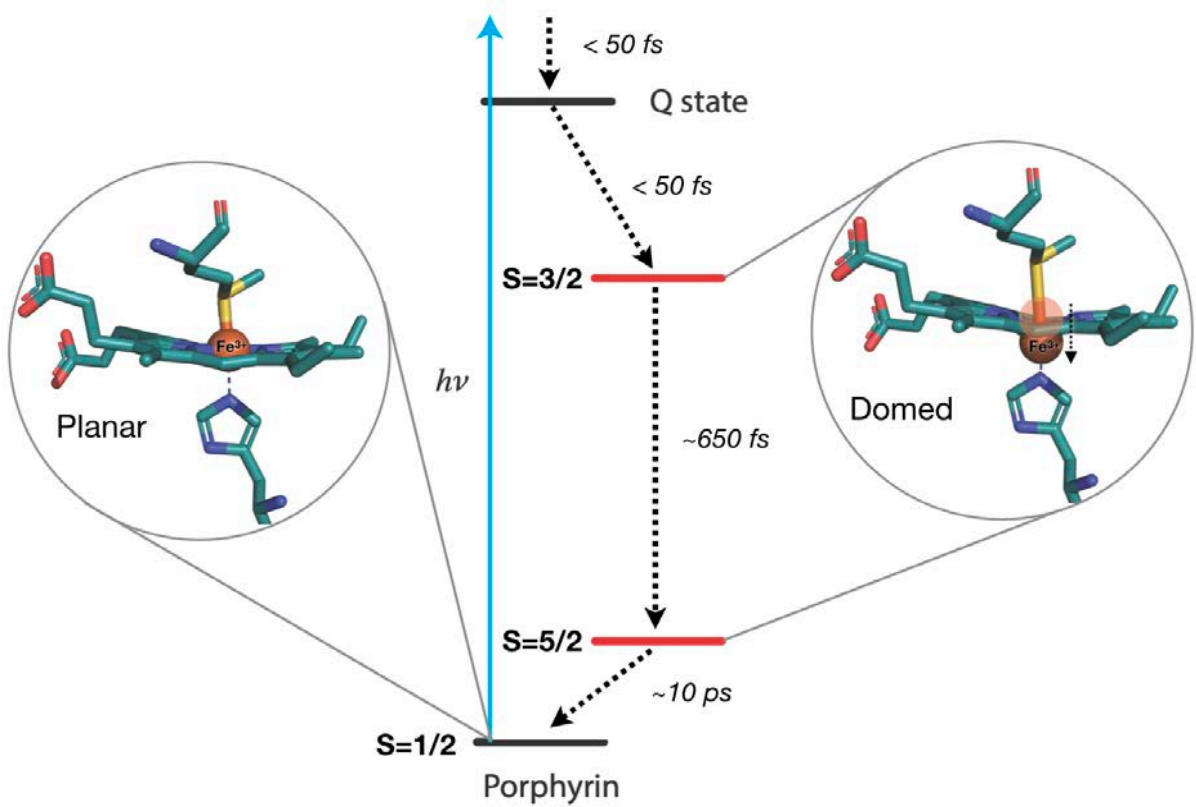

**Figure 6**